# Comparison of Proposed Data Dissemination Protocols for Sensor Networks Using J-Sim

[1]Deepali Virmani
Assistant Professor IT Dept.
GPMCE, IP University.
Delhi, INDIA.
deepalivirmani@gmail.com

[2]Satbir Jain
Assistant Professor CSE Dept.
NSIT, Delhi University.
Delhi, INDIA.

***Abstract*-A distinguishing characteristic of wireless sensor networks is the opportunity to exploit characteristics of the application at lower layers. This paper reports on the results of a simulation comparison of proposed data dissemination protocols using the J-Sim simulator for the WSN protocols: Forwarding Diffusion Data Dissemination(FDDDP) , Decentralized Data Dissemination(DDDP), Credit Broadcast Data Dissemination (CBDDP*), *Energy Aware & Geographical Data Dissemination (EAGDDP) .Our performance provides useful insights for the network designer such as which protocols (and design choices) scale control traffic well, improve data delivery or reduce overall energy consumption ,improves routing overhead and maximizes the bandwidth utilization. The static pre configuration of the cell size in DDDP, is one of the reasons why DDDP exhibits larger routing overhead than FDDDP by 74.2% on average. Although CBDDP produces approximately 94.6% smaller overhead than DDDP and 90.7% smaller than FDDDP, because of statically configured amount credit CBDDP delivers on average 7.5 times more of the redundant data packets than DDDP and FDDDP.EAGDDP improves the delivery by 80% on average and makes a balance of energy consumption .We suggest that making these protocols truly self-learning can significantly improve their performance.**

## I. INTRODUCTION

The rapid advances in wireless communication and Micro Electro Mechanical System (MEMS) have made Wireless Sensor Networks (WSNs) possible. Such environments are typically comprised of a large number of sensors being randomly and densely deployed for detecting and monitoring tasks. These sensors, developed at a low cost and in small size (mm-scale for smart dust motes [1]), are responsible for object sensing, data processing, storing, and routing activities. Applications of such networks range from battlefield communication systems (e.g. intrusion detections and target surveillance) to environmental monitoring networks such as habitat monitoring, chemical sensing, infrastructure security, inventory and traffic control etc. For example, sensors are distributed across a forest in order to report the origin of a fire event when there is a significant increase in the average monitoring temperature. Reference [2] provides a more thorough discussion on some potential WSN applications. Unlike the conventional adhoc communication networks, energy resources in WSNs are usually scarce due to the cost and size constraints of sensor nodes. In addition, it is impractical to replenish energy by replacing batteries on these nodes. Conserving energy is thus the key to the design of an efficient WSN. WSNs may deploy several hundreds to thousands of sensor nodes. Protocols in such networks must therefore be scalable. Furthermore, since nodes are dynamic and their geographic positions are not pre-determined, these nodes may also need to possess some self organizing capabilities. Network dynamics that result from both node movement and unpredictable energy depletion also bring new challenges to the design of an efficient WSN. Since nodes can only carry limited battery resources, they usually get disconnected from the network easily. Such frequent node disconnections suggest that the design must accommodate topology changes. Communication in wireless sensor networks is data-centric and must minimize the energy consumed by unattended battery-powered sensor nodes [3][4][5][7] . Our key observation is that despite their design intentions to make these protocols self-configuring, they in fact rely on a significant number of statically configured parameters. We suggest which parameters for each protocol should be dynamically configured in response to measured network state, using passive measurement techniques such as Bayesian inference to reduce the measurement overhead. Making these protocols truly self-learning techniques could significantly improve their performance.

Section II describes proposed data dissemination protocols, Section III gives the methodology of simulation , Section IV describes overview of various metrics used for comparison, Section V shows the result of comparison and section VI gives the summary and suggestions for future work.

## II. PROPOSED DATA DISSEMINATION PROTOCOLS :INTRODUCTION

### *A.. Forwarding Diffusion Data Dissemination Protocol(FDDDP)*

FDDDP (Fig.1) is the first proposed data centric communication protocol for wireless sensor scenarios. The data generated by the source node is named using attribute value pairs. The consumer node requests the data by periodically broadcasting an request for the specific data. Each node in the network will establish a link towards its neighboring nodes from which it receives the request. The link specifies both the data rate and the direction towards which the data should be sent. Once the source node detects an interest it will send exploratory packets towards the consumer, possibly along multiple paths. As soon as the consumer begins receiving exploratory packets from the source it will select one particular neighbor from whom it chooses to receive the rest of the data. The data will then flow back towards the consumer along the selected path. The selected path packets are also used for local path repairs in case of the failure of some nodes during the data delivery phase.

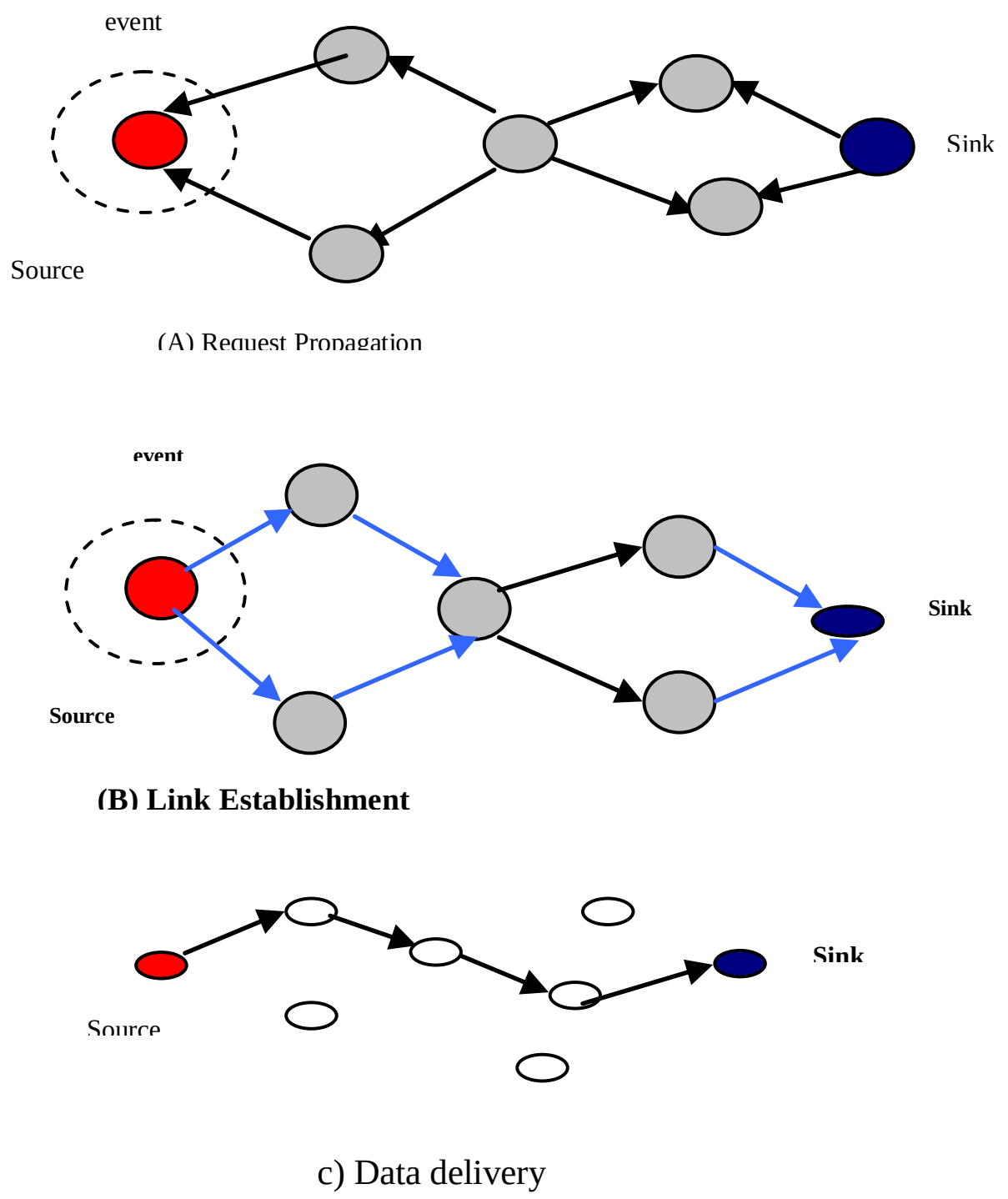

Fig. 1 FDDDP

### B. *Decentralized Data Dissemination protocol(DDDP)*

DDDP (Fig. 2) is based on decentralized architecture. It uses a cell like structure to divide the entire topology into small cells. Only sensors located at a cell boundary need to forward the data. The consumer actively builds this cell structure through the network and sets up forwarding points in the sensors closest to the cell boundary called centralized nodes (CN). One level is the cell at the consumer's current location and the other one is the CN at cells boundaries. The consumer only floods the query within its own cell. When the nearest CN that hears the query, it forwards it to its adjacent CNs(of the same cell or the next cell ). This process continues until the query reaches the producer or one of the CNs that have the corresponding data. During the query propagation period the network establishes the reverse path towards the consumer for the reply, so that it can enable the data path to be the same as that of the query propagation.

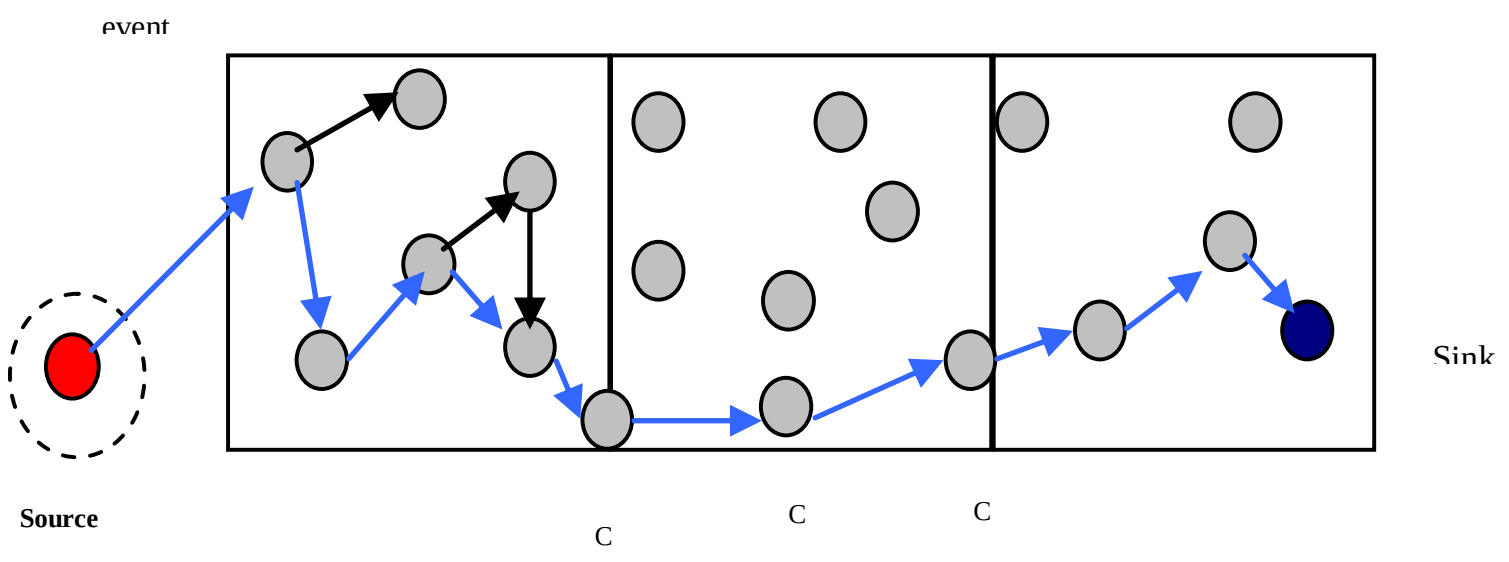

Fig. 2 DDDP

### C. *Credit Broadcast Data Dissemination Protocol (CBDDP)*

In CBDDP (Fig. 3) a node on deployment sets its cost to reach the consumer at infinity. As soon as the consumer node starts up it broadcasts the advertisement message containing its initial cost. Each intermediate node that hears the advertisement will calculate the receiving cost of the message. At the end of the cost field setup period each working node will have calculated the minimum cost for it to reach the consumer. Each message carries a ”credit” in its header in addition to its optimal shortest path cost for transmission. Depending on the ”credit” amount data packets can flow along multiple paths rather than a single optimal shortest path .The packet will eventually arrive at the sink node through at least one of the working paths even if some intermediate nodes malfunction or if channel gets corrupted . If the ”credit ” is set to be higher that the minimum cost. Each intermediate node will make its own decision regarding the forwarding of a packet based on the amount of credit in the data message, its own minimum cost value and the remaining ratio. CBDDP assumes a static network so node movement will require excessive updates of cost field. Each data packet will carry in its header the minimum cost of the source node to reach to the consumer ($Cost_{source}$) , some constant ($\beta$), the current energy used ($E_{current}$) and the sender's minimum energy ($E_{min}$) . The Remaining Ratio is calculated as follows.

Let .

RR - Remaining Ratio

Th - Threshold Value

If RR is bigger than Th then a node will rebroadcast the message.

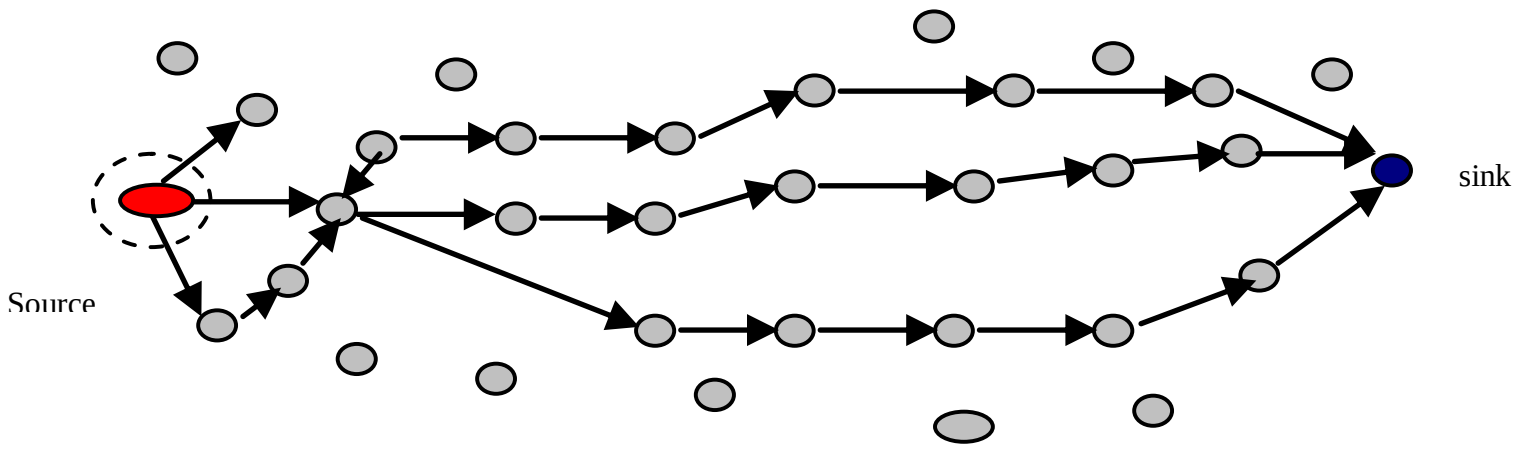

Fig. 3 CBDDP

*D. Energy Aware &Geographical Data Dissemination Protocol (EAGDDP)*

In EAGDDP (Fig.4) takes residual energy into consideration and is designed to efficiently disseminate queries to a destination. As queries are often geographical (i.e. they have a target area), packets are directly forwarded to the particular destination rather than flooded everywhere. EAGDDP assumes that nodes are aware of their own geographic positions, and uses energy-aware neighbor selection to aggressively route the queries toward the specific target region [8][9]. In addition to the distance to destination, neighbor's residual energy is also considered in the cost function so that energy load among any neighborhood can be balanced. The tradeoff, however, is the increased path length used to transmit the queries since energy efficient paths are not necessarily the shortest. Restricted forwarding immediately follows to disseminate packets inside the area once the queries have arrived at the border of the region. In this protocol we assume that the node N is forwarding packet P whose target region is R . The centroid of the target region is D. Upon receiving the packet P, the node N routes P progressively towards the target region and at the same time tries to balance the energy consumption across all the neighbors. Node N achieves this trade off by minimizing the learned cost l(Ni,R) value to its neighbor Ni. Each node N maintains l(N,R) which we term as learned cost to the region R. A node frequently updates its l(N,R) value to its neighbors. If the node does not have l(Ni,R) state for a neighbor Ni, it computes estimated cost e(Ni,R) as the default cost for l(N,R).

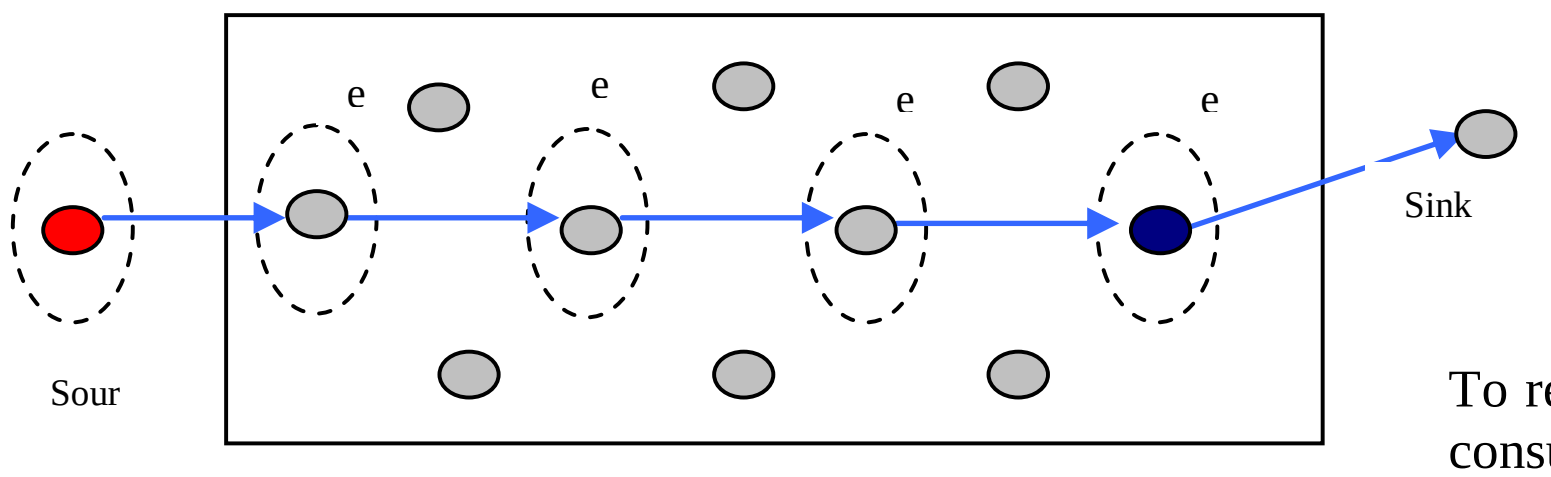

Fig. 4 EAGDDP

Estimated cost e(Ni,R) of Ni is calculated as

$$e(Ni,R) = \mu d(Ni,R)+(1-\mu)e_c(Ni) \quad (1)$$

$\mu$ - tunable weight
d(Ni,R) - distance from Ni to centroid d to region R
$e_c$ - consumed energy

as soon as nodes picks up a next node-hop neighbor it sets its own l(N,R) to l(Nmin,R) + C(N,Nmin) which is known as cost of transmission from N to Nmin .

## III. METHODOLOGY

This section describes the simulation methodology and the metrics used for the comparison of protocols. J-Sim (Java Simulator for sensor networks ) was used for the simulation of protocols. Each of the data dissemination protocols studied has the same underlying IEEE 802.11 MAC layer, the same radio propagation model based on the 954Mhz frequency of the Lucent WaveLan DSSS radio with omni-directional antenna placed 1.9 meters above the node and the same data load. 2 different topologies with uniformly distributed nodes have been generated. The size of the topology, the number of nodes that are deployed and the SNINDA (the specific Number of nodes In Nominal distance Area) can have significant impact on protocol behavior. The same topology scenarios are used across different protocol simulations. Given the radio range of a node, the topology size and the number of nodes deployed (SNINDA) represents the largest possible number of neighbors that a node can hear from and is calculated according to following formula

S - number of nodes (size of a topology)
a - area of the topology
r - radius range of radius

Table I. Shows the parameters used for generating the various simulation topologies.

TABLE I. Parameters used for generating the various simulation topologies

| Number of nodes | Dimensions | SNINDA |
|---|---|---|
| 20 | 340 x 340 | 40 |
| 40 | 511 x 511 | 40 |
| 60 | 626 x 626 | 40 |
| 80 | 713 x 713 | 40 |
| 100 | 810 x 810 | 40 |
| 120 | 886 x 886 | 40 |
| 140 | 911 x 911 | 40 |
| 160 | 994 x 994 | 40 |

To represent the worst case scenario only one source and one consumer used for each simulation. The source and consumer are located at opposite sides of the topology so that a large number of Eight different topology scenarios are used for the simulation. The first one consists of 20 nodes in the topology. The number of nodes deployed is progressively increased by 20 until there are 160 nodes in the topology. Data packets are generated at intervals of 2 second. The simulation is run for 500 seconds therefore each protocol has enough time to discover the route from the consumer to the producer and produce substantial amount of data traffic.

## IV. METRICS

For the evaluation of protocols the following four metrics have been chosen. Each metric is evaluated as a function of the topology size, the number of nodes deployed, the SNINDA and the data load of the network.

### A. Average Energy Consumption ($E_{avg}$)

The average energy consumption is calculated across the entire topology [10]. It measures the average difference between the initial level of energy and the final level of energy that is left in each node. This metric is important because the energy level that a network uses is proportional to the network's lifetime. The lower the energy consumption the longer is the network's lifespan.

### B. Routing Overhead

This metric represents the total amount of routing packets transmitted during the simulation time. Let

Tp = the total amount of routing packets that a node transmits during the simulations

n = the number of nodes deployed

Then
$$R_{OH} = \sum_{k=1}^{n} T_{PrK} \quad (2)$$

This metric is important for the comparison of these protocols as it indicates the scalability of a protocol. Each protocol has to function in low bandwidth and congested environments, so this metric is a good indication of the degree of functionality for a protocol and its efficiency in terms of resources consumption. Also it operates as a very good indication of how much effort is needed to construct and maintain a route between the source and the consumer.

### C. Packet Delivery ratio (Dr)

This metric represents the ratio between the number of data packets that are sent by the producer and the number of data packets that are received by the consumer. Let

Psent = the number of packets sent by the source

Prec = the number of packets received by the consumer(including duplicates)

Then

$$Dr = \frac{P_{sent}}{P_{rec}} \quad (3)$$

This metric indicates both the loss ratio of the routing protocol and the effort required to receive data. In the ideal scenario the ratio should be equal to 1. If the ratio falls significantly below the ideal ratio, then it could be an indication of some faults in the protocol design. However, if the ratio is higher than the ideal ratio, then it is an indication that the consumer receives a data packet more than once. It is not desirable because reception of duplicate packets consumes the network's valuable resources. The relative number of duplicates received by the consumer also important because based on that number the consumer, can possibly take an appropriate action to reduce the redundancy.

### D. Bandwidth utilization

Bandwidth is defined as the amount of total aggregated data at the nodes, which is transferred through the links [6]. Bandwidth utilization is calculated as the maximum amount of data that is passed through the links in full duplex mode. Measurements are based on sampling intervals taken in the network. Let

$\Delta$ = full cycle (full duplex mode)

$\Delta_{in}$ = bandwidth required for aggregating data at the nodes

$\Delta_{out}$ = bandwidth required for transmitting the data in network

$N_{speed}$ = Network Speed

8 = Sampling rate

100 = time interval between two samples(nsec)

$$Band_{util} = \frac{\max(\Delta_{in}, \Delta_{out}) \times 8 \times 100}{(No\ of\ \sec onds\ in\ \Delta) \times N_{speed}} \quad (4)$$

This is important because more the bandwidth utilization less will be the energy consumption and less number of bytes will be wasted in network configuration and congestion. As the result of this metrics battery life will be improved.

## V. COMPARISON RESULTS

### A. Average Energy Consumption

Fig. 5 shows the relative energy consumption of all four protocols. As expected CBDDP shows the highest energy consumption in comparison as compared to DDDP and FDDDP .EAGDDP also shows higher average energy consumption. DDDP and FDDDP have very similar energy consumption with DDDP being slightly higher from 20 to 80 and at 160 nodes. DDDP also performed marginally better for the 120 nodes scenario. The reason being limited flooding of packets to one cell only. Therefore choice of cell size is important in DDDP .Table II summarizes the cell size used for DDDP based on number of nodes deployed and default size of the topology. If the number of cells is small then DDDP will flood its data similarly to FDDDP. As the no of cells grow the flooding is constrained to an area in network. That's why there is increase in energy consumption .The ideal cell size is not investigated in this work.

TABLE.II Shows the variations in the cell size

| Number of nodes | Dimensions | No of cells |
|---|---|---|
| 20 | 340 x 340 | 4 |
| 40 | 511 x 511 | 9 |
| 60 | 626 x 626 | 12 |
| 80 | 713 x 713 | 20 |
| 100 | 810 x 810 | 23 |
| 120 | 886 x 886 | 28 |
| 140 | 911 x 911 | 32 |
| 160 | 994 x 994 | 37 |

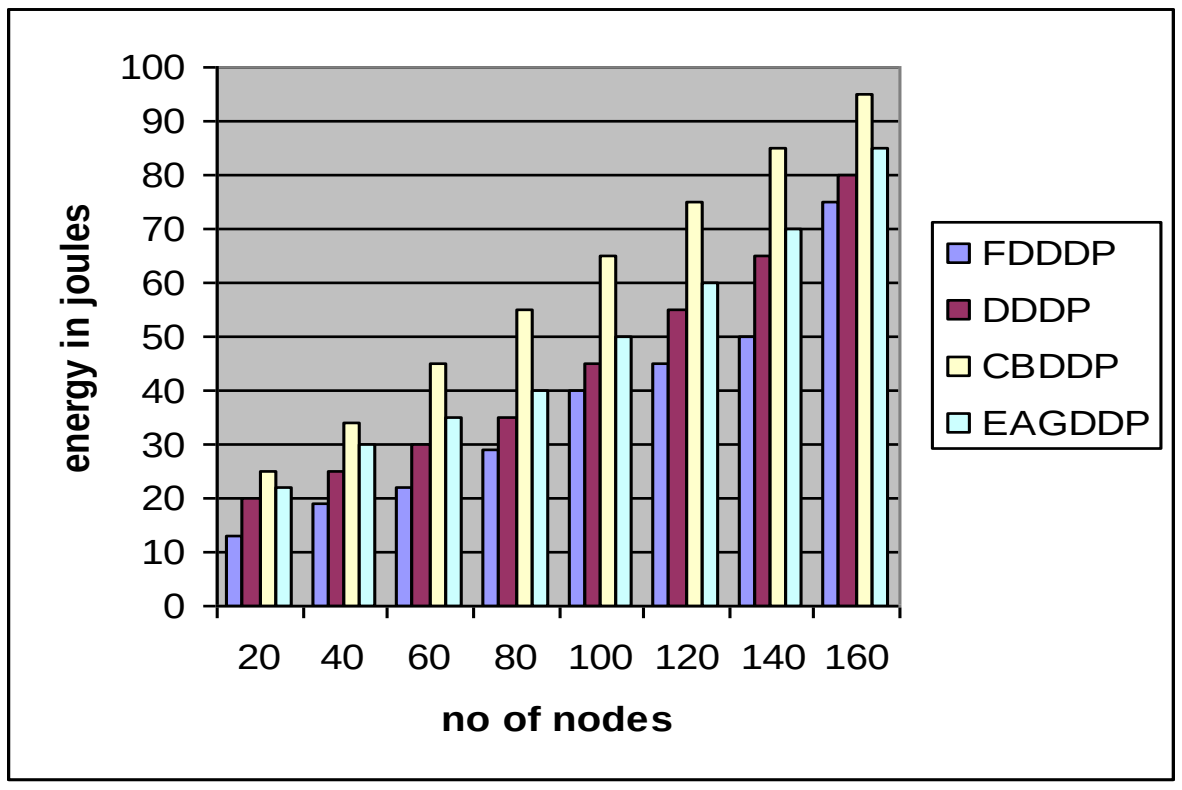

Fig. 5 Average Energy Consumption

*B. Routing Overhead*

Fig. 6 shows the relative routing overhead for all four protocols .As can be seen ,DDDP exhibits the largest routing overhead .This is the indication that cell like structure is very expensive for DDDP in terms of routing overhead . additionally, the size of the cell plays a vital role in the behavior of DDDP. The cell size has to be set before the simulation stars and there is no way to change it in order to respond to changers in environment.

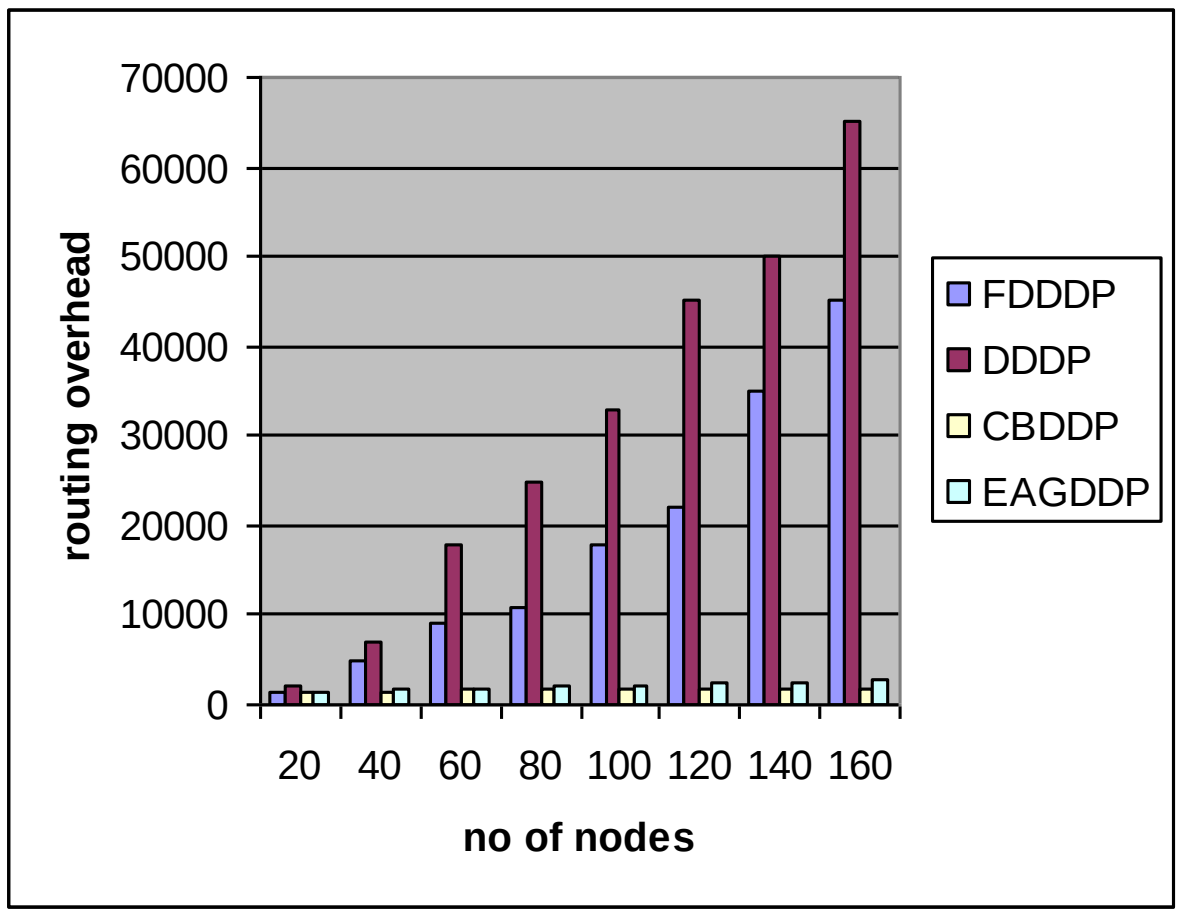

Fig. 6 Routing Overhead

Although as shown as in Fig.7 the number of routing overhead packets produced by CBDDP and EAGDDP Fluctuates significantly across simulations and therefore they have most unpredictable behavior in terms of routing overhead .Overall CBDDP has the smallest routing overhead .The refreshment of cost field in response to major changes appears to be a very positive feature for CBDDP and EAGDDP.

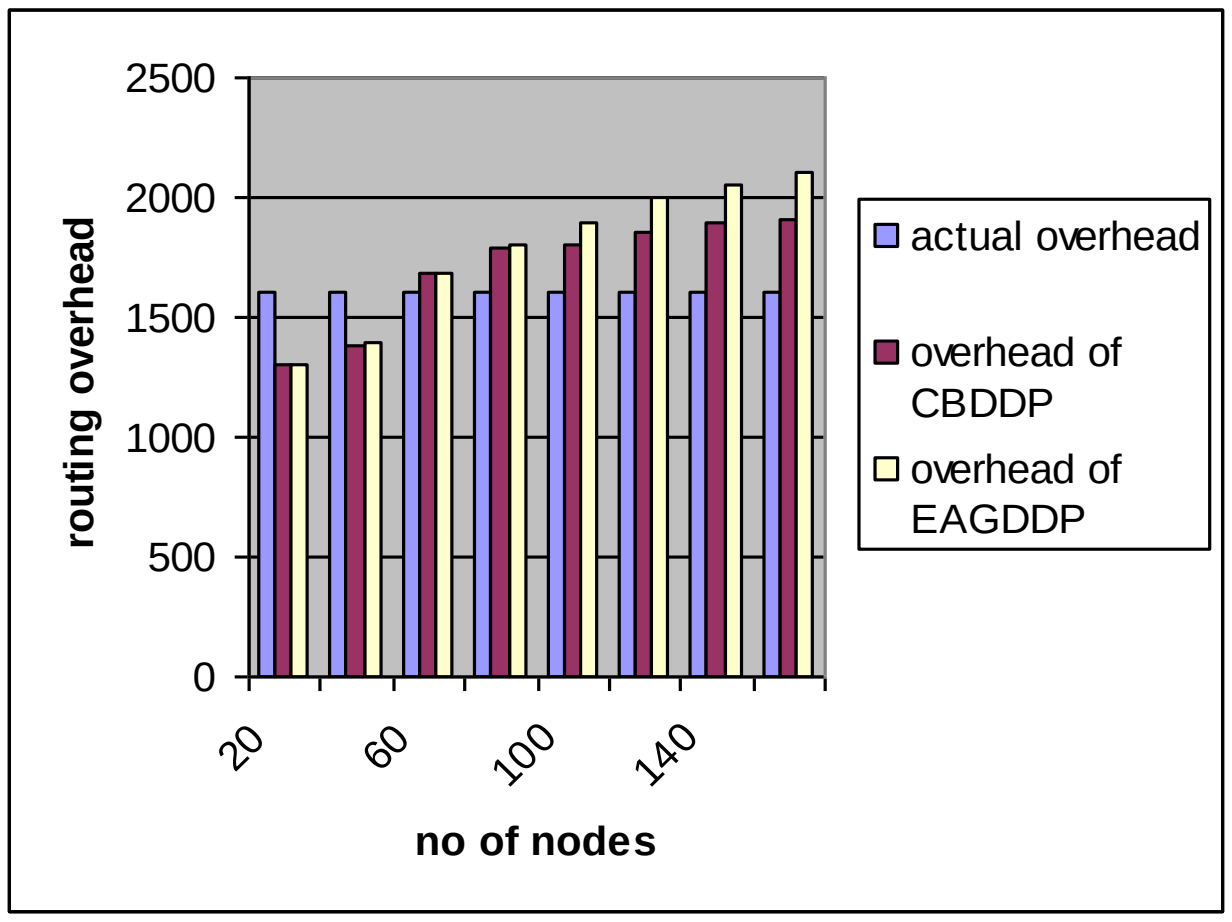

Fig. 7 CBDDP & EAGDDP Overhead

*C. Delivery Ratio*

Fig. 8 shows the relative delivery ratio of data packets for all the protocols. DDDP and FDDDP have very similar delivery ratios and very close to the ideal one. FDDDP, however has slightly more fluctuations. EAGDDP has the highest delivery ratio, CBDDP on the other hand has a larger delivery ratio than the other two protocols with a very large error bars. Therefore even for the constant amount of the credit and the stable topology of nodes we can not predict the exact delivery for CBDDP at the beginning. It is also much higher than the ideal one. This feature of EAGDDP and CBDDP may increase the robustness of data delivery in the case of noisy channels. However, this feature is not particularly desirable while operating on clear channels, as it leads to high energy consumption.

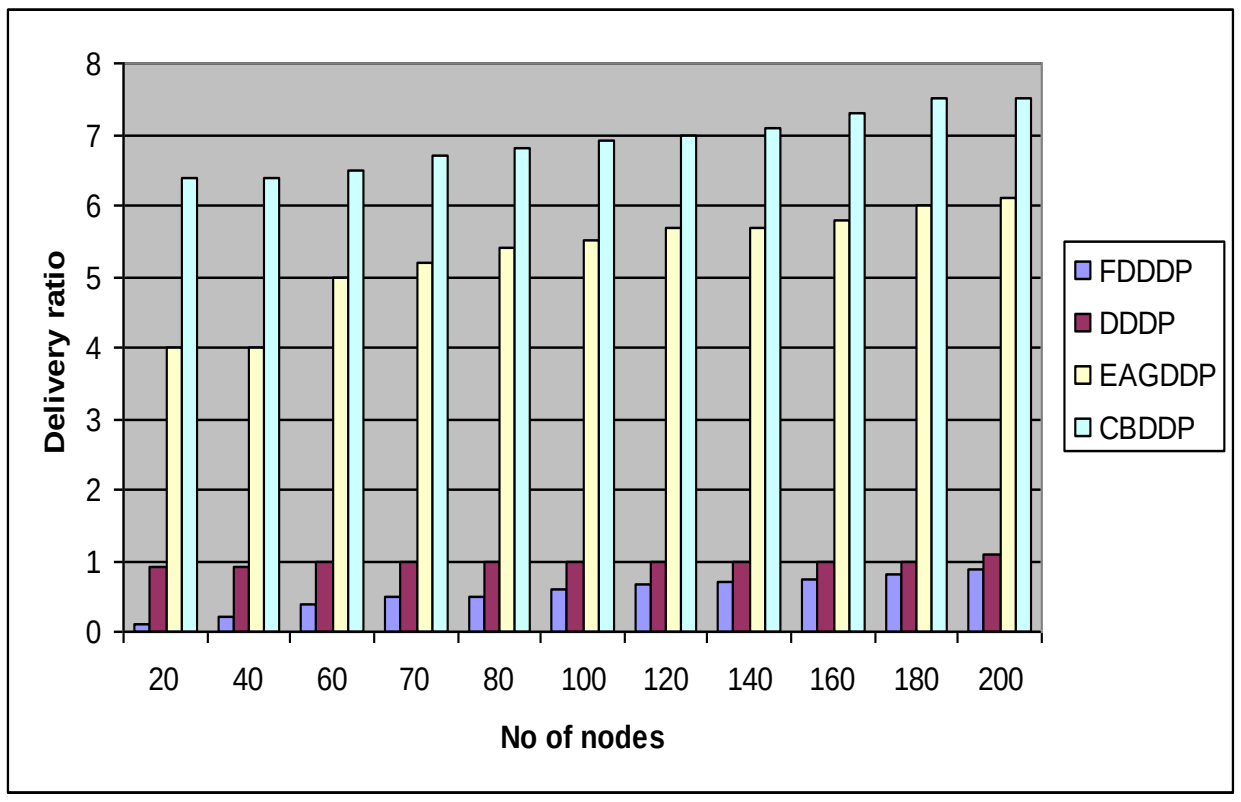

Fig. 8 Delivery Ratio

### *D. Bandwidth Utilization*

Fig. 9 shows the no of bytes used per sec for all the protocols .EGADDP shows the best results by wasting minimum number of bytes .As 500 bytes/sec are transmitted for all the protocols.

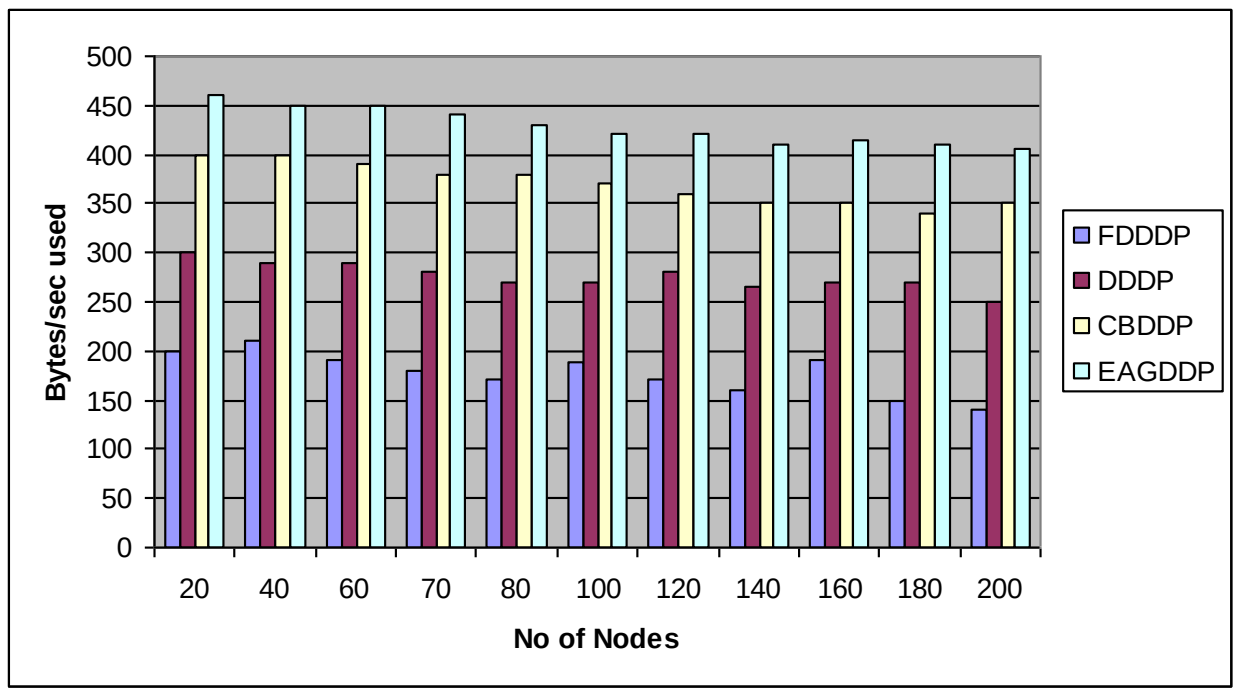

Fig. 9 Bytes utilized per sec

## VI. CONCLUSIONS, SUGGESTIONS AND FUTURE WORK

### *A. Conclusion*

This paper presented the comparison between four data dissemination protocols (FDDDP, DDDP, CBDDP, EAGDDP), for wireless sensor networks, using J-Sim simulations .These protocols cover a large number of design choices including the construction of the cell structure, credit-based adjustable mesh forwarding and the establishment of links for neighboring nodes and finding geographical position of neighbors . Typically, when these protocols are simulated in isolation, the emphasis is on studying only the scaling behavior of the protocol (for example, the impact of network density on scaling behavior). Such an approach can mask the design weaknesses of a particular protocol. Being a relative performance comparison, these simulations provide useful insights to what kind of design choices are the most desirable in order to improve the performance of proposed protocols. Each of the protocols performed well in some cases, but displayed certain drawbacks in others. The performance of DDDP and FDDDP was quite close, where as performance of CBDDP and EAGDDP was very close. But the performance of later two is quite distinctive from the previous two.

DDDP has 74.6% large routing overhead but consumes only 3.9% more energy than FDDDP. This is due to the nature of data forwarding in DDDP. It constrains the flooding of data packets to one cell. However, for large cell sizes relative to the topology size it floods the data in a very similar way to the flooding of interests and exploratory packets used by FDDDP.

TABLE III. Summary of protocols and Suggestions for improvement

| Parameters | FDDDP | DDDP | CBDDP | EAGDDP |
|---|---|---|---|---|
| Routing Overhead | Low | Worst | Best | Equal to CBDDP |
| Data Delivery Ratio | 0.4---0.9 | Ideal(1) | Highly Redundant | Best |
| Energy Consumption | Lowest | Average | Worst | Equal to CBDDP |
| Bandwidth Utilization | Worst | Average | High | Best |
| Suggestions | Minimize forwarding to all paths by adapting refresh rates according to measured path latencies to improve bandwidth utilization | Improve routing overhead by adapting cell size according to measured network density | Reduce Redundancy by adapting credit according to measured path loss | Overall parameters are satisfactory but energy consumption is to be minimized to achieve best results by minimizing the path length |

CBDDP has 90.7% smaller routing overhead than FDDDP and 94.6% smaller than DDDP because of the way it refreshes its minimum cost at each node. The cost is refreshed only when there are major changes in the network topology are detected or the delivery of the data has been delayed. However, because of the way it forwards its data to the consumer it consumes redundantly 32% large amount of energy compared to FDDDP and 23.7% larger compared to DDDP. Overall FDDDP consumes 4.5% less amount of energy than DDDP.

EAGDDP uses energy aware and geographically informed neighbor selection to route a packet towards the target region. This strategy attempts to balance energy consumption and thereby increase network lifetime. Within a region, it uses a recursive geographic forwarding technique to disseminate the packet .simulations show that that Delivery ratio is high in EAGDDP .EAGDDP delivers 82% more packets as compared to DDDP and 79% more packets than FDDDP.,EAGDDP delivers 20% more packets in non uniform traffic .Routing overhead is also decreased to 70% as compared to DDDP. Bandwidth utilization is highest in EGADDP but energy consumption is also maximum Finally, DDDP has a slightly closer delivery ratio to the ideal ratio than does FDDDP, although the delivery ratios are very similar in both of these protocols. FDDDP appears to have larger fluctuations for the delivery ratio of data packets. The smallest ratio of data packets delivery was approximately 0.75 whereas the DDDP delivery ratio did not fall below 0.9 during the simulation period.

*B. Suggestions*

Comparisons revealed that the performance of a protocols was enhanced where its parameters was not inflexibly predetermined but rather, could be varied by adapting to its environment. To boost their performance, we suggest that making these protocols truly self-learning by configuring protocol parameters in response to measured network state, using passive measurement techniques such as Bayesian inference. DDDP has static cell size performance would be improved if it could adjust its cell size according to environment inorder to limit flooding .CBDDP can be improved by adding the ability of its consumers to adjust the credits that packet carries in order to reduce redundancy. This credit could be the function of the application reliability requirements and dynamically configured as a function of mean percentage packet loss along a given path which can dynamically derived based on statistics. EGADDP can be improved by minimizing path length. We suggest that parameters for each protocol such as credit (CBDDP), cell size(DDDP), refresh rate (FDDDP) and the geographical positions (EAGDDP) should be dynamically configured in response to measured network state, such as path loss, latency, network density and diameter, using passive measurement techniques such as Bayesian inference. In summary, making these protocols truly self-learning could significantly improve their performance. Table III summaries the results and suggestions of our comparison